\documentclass[fleqn,10pt]{article}
\usepackage[utf8]{inputenc}
\usepackage[T1]{fontenc}
\usepackage{graphicx} 
\usepackage{hyperref}
\usepackage{lineno,color}
\usepackage{amsmath,amsfonts,amssymb}
\usepackage{authblk}
\usepackage{geometry}

\date{}
\title{Universal Database for Economic Complexity}

\author[1,2,*]{Aurelio Patelli}
\author[2,1]{Andrea Zaccaria}
\author[1,3]{Luciano Pietronero}
\affil[1]{Centro di Ricerca Enrico Fermi, Via Panisperna 89 A, I-00184 Rome, Italy}
\affil[2]{Istituto dei Sistemi Complessi (ISC) - CNR, UoS Sapienza,P.le A. Moro, 2, I-00185 Rome, Italy}
\affil[3]{Dipartimento di Fisica Universit\`a “Sapienza”, P.le A. Moro, 2, I-00185 Rome, Italy}

\affil[*]{corresponding author: Aurelio Patelli (aurelio.patelli@cref.it)}

\begin{document}

\flushbottom
\maketitle

\begin{abstract}
	We present an integrated database suitable for the investigations of the Economic development of countries by using the Economic Fitness and Complexity framework.
	Firstly, we implement machine learning techniques to reconstruct the database of Trade of Services and we integrate it with the database of the Trade of the physical Goods, generating a complete view of the International Trade and denoted the Universal database. 
	Using this data, we derive a statistically significant network of interaction of the Economic activities, where preferred paths of development and clusters of High-Tech industries naturally emerge. 
	Finally, we compute the Economic Fitness, an algorithmic assessment of the competitiveness of countries, removing the unexpected misbehaviour of Economies under-represented by the sole consideration of the Trade of the physical Goods.
\end{abstract}

\section*{Introduction}
Economic Fitness and Complexity~\cite{Tacchella2012,Tacchella2013,Cristelli2013,Caldarelli2012,zaccaria2016case,Sbardella2018} (EFC) is a novel conceptual and practical framework for the estimation of the competitiveness of nations and the relatedness between sectors, borrowing concepts and methods from Statistical Physics and the Complex Systems Science~\cite{Pietronero2008}.
A key feature of the method resides in its bottom-up, data-driven approach, which relies on methods like complex networks and machine learning to reconstruct and investigate the ecosystem constituted by the economic actors and their activities; moreover, all scenarios are based on empirical observations and certifiable hypotheses.
This approach departs from the canonical econometric narrative, where the economic performances are usually gauged by monetary metrics, such as the Gross Domestic Production (GDP).
Instead, EFC aims to capture the competitiveness of a country, and not its wealth, by introducing a synthetic and non-monetary metric, the Fitness~\cite{Tacchella2012}, bringing to light new and relevant economic patterns.

A key feature of Economic Fitness and Complexity is its reliance on homogeneous and high-quality data from which it effectively extracts information by aiming to achieve a maximal signal-to-noise ratio.
In the \emph{Big-Data} era, the novel and great size of available databases may generate a broad confidence on the new possibilities offered by large scale analyses, although issues related to the quality of the gathered data and the specific investigations are sometime underplayed~\cite{Hosni2018}.
It is rather intuitive that Big-Data may present Big-Noise~\cite{Silver2012}, and so a careful selection of the sources must be accomplished at the starting point of the research in any data-driven field. 
The reference database in the classical EFC analysis is the International Trade database reconciled and regularized by Tacchella and collaborators~\cite{Tacchella2018} starting from the UN-COMTRADE data. 
This database covers the external flows of physical Goods between the countries in the World with a very fine level of detail of the classification collecting about 5000 product's codes. 
EFC is based on the export database for various reasons, both practical, due to its homogeneity and standardization across different countries, and conceptual. 
The idea is to deduce the competitiveness of countries not explicitly looking at their capabilities, but by inferring the presence of such endowments in a indirect way, that is, by looking at the final outputs, the products that such countries are able to export~\cite{hausmann2007you}. 
Not trivially, the GDP predictions obtained by EFC and based on the Trades indicate a quantitatively better performance compared to the best conventional economic models~\cite{Tacchella2018} and with a much lower data requirement.

Unfortunately, the Services are not included in the set of available features in the UN-COMTRADE and, albeit available from an IMF database (www.data.imf.org), neither in the usual analyses of EFC, despite a relevant and growing fraction of money flows through the channel of the Services~\cite{loungani2017world}.
A first attempt to complement the Services into the aforementioned analysis was followed by~\cite{Stojkoski2016}, finding that at a very aggregated scale the Services tend to be more complex with respect to the Goods.
However, the authors complain that the high level of aggregation may be the cause of a loss of information necessary to accurately separate the Complexity of the Activities and the Fitness of the Countries.
With a deeper level of aggregation, the authors in ~\cite{Zaccaria2018services} show a more heterogeneous scenario of the Complexity of the Activities, pointing out that complex Services cluster with complex Goods.
In the EFC narrative the clustering of a set of Activities is the signal indicating that there is a participation of common intangible capabilities~\cite{Cristelli2013}, necessary to be competitive in both the domains.
However, either manuscripts select a subset of the available economies in the World because the covering of the databases implemented in the analysis present many missing features.
Indeed, both the references consider only 116 countries, with important missing economies such as China and Great Britain. 
Recently, the work of Mishra and collaborators~\cite{Mishra2020} shows the importance to aggregate the Services and the Goods in a common database especially accounting properly the economic relevance of the developing nations. 
Saltarelli et al.~\cite{saltarelli2020export} use the World Input-Output Database (WIOD) to show that export mirrors remarkably well domestic production for manufacturing sectors, but this relation fades away for service related sectors. 
All these analyses, however, use a limited number of services sectors because of the high number of missing values present in the original IMF database; a limitation that the present work addresses.

The analysis endowed by the previous works highlight the purposes of the Services in the correct estimation of the competitiveness of the nations.
However, the reference database for the Services, the \textit{Balance of Payments and International Investment} (BOP) data collected by the International Monetary Fund (IMF)~\cite{BOPS}, presents some weakness and important missing scores, especially in comparison with the quality of the Goods database. 
Strikingly, the actual IMF-BOP classification is unsuitable for the EFC analysis, presenting an overlapping and convoluted hierarchy of sector.
The first core result of the present paper is the reconciliation of the quality of two databases, the Goods and the Services, obtained by reclassifying the IMF services sectors and reconstructing the missing elements of BOP by using and comparing different machine learning techniques.
The best reconstruction method obtained is then used to create the so-called \textit{Universal} database of EFC, aggregating Services and Goods for 160 countries and a total of 124 sectors, providing the largest set of common nations and sectors available from both the BOP and the Trade datasets.
Finally, the Universal database is used in the Economic Fitness and Complexity analysis, obtaining two conceptual and practical advances: i) a statistically validated network of universal sectors, that will be of practical use to predict and recommend new sectors of development, and ii) the computation of the Universal fitness, the novel EFC indicator to assess the competitiveness of nations, now including also the services.

\section*{Methods}
This section introduces the database used in the later analysis, aggregating Goods and Services; in particular, we will focus on the motivations and the methodologies we adopted for the re-classification of services and the reconstruction of the missing values.

\subsection*{Goods: the International Trade}
Following the definition of the IMF~\cite{BOPS}, a Good is a physical item or commodity over which ownership can be passed via transaction. 
Consistently, it is recorded by the customs and available via UN-COMTRADE.
Following the standard approach of EFC, the reference database collecting the scores of the Goods is the reconciliation of the International Trade database contained in \textit{UN-COMTRADE}~\cite{COMTRADE}.
It collects the export flows of classes of physical products between nations.
The classification of the product follows the Harmonized System revised in 1992 (HS92), consisting of 5040 sectors labelled by 6-digits codes at the sub-heading level.
The temporal coverage of the database runs from 1996 to 2018, spanning slightly more than two decades, and it is made available for 169 countries corresponding to the principal economic players in the World.

The statistic of the codes at 6-digits does not present a level of details and a statistics comparable to the orders of magnitude of the data of the Services.
It is also worthy to note that the HS classification varies in time and in the available years a relevant modification happens in 2007.
Perhaps, such modification induced a re-balance in some sectors at 6-digits that create a non-uniform transition in few and low complexity codes~\cite{JRC_report}.
Therefore, we aggregate the classification at the closest comparison to the Services database, being at the 2-digits classification. Note that this choice also mirrors the respective weight in the international trade \cite{loungani2017world}.
Furthermore, the 2-digits classification has 97 codes, but we remove one sector (the code 99, the \emph{Commodities not specified according to kind}) because it does not represent a well defined sector and has a low impact on the overall exports.


\subsection*{Services: description and classification issues}
According the IMF~\cite{BOPS}, a Service is the result of a production activity, or facilitate the exchange of Goods, or is a financial asset.
Therefore, Services are usually non-separable items and cannot be separated from their production.
The Service database is based on the \textit{Balance of Payments and International Investment} (BOP) data collected by the International Monetary Fund (IMF)~\cite{BOPS}~\footnote{The data has been downloaded the 28th of January 202 from the website www.data.imf.org}.
The BOP database offers three measures of Services: the credit, the debit and the net values. In comparison with the measure available for the Goods, we select the credit indices, since they correspond to exported services. Two closely related classifications are present: the full BOP classification implemented by the IMF and an \emph{alternative} alpha-numerical classification reported in their metadata.
The full BOP classification is based on the composite nature of the codes of the Services but it presents a key drawback for the analytical purposes accounted below.
In fact, it is possible that the sum of the sons of a code have a total value larger that the assigned value of the parent.
This misleading situation happens because the classification may implement different methodologies in order to generate the hierarchy.

\begin{figure}[!t]
	\centering
	\includegraphics[scale=0.4]{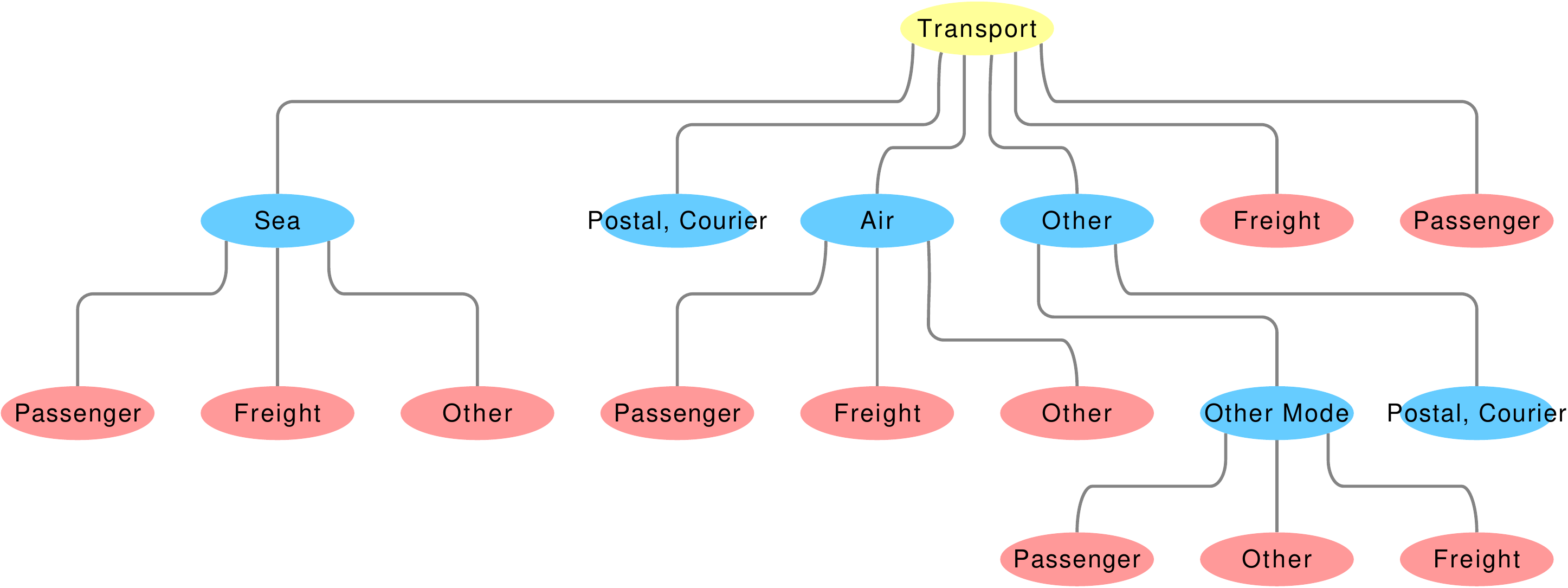}
	\caption{The network of the transport and its sons divided with the two possible dis-aggregation of the codes. Only the blue codes sum up to the value of the parent. The name of the nodes corresponds to the last part of the definition available from the BOP's metadata and the full definition correspond to the sequential aggregation of the definitions of the parent nodes.}
	\label{fig:network_transport}
\end{figure}
In order to explain this situation we discuss an example regarding the case of the  sons of the `\emph{Transport Services}' (`STR' codes) where the dis-aggregation considers both the kind of transportation implemented or the nature of the transported Goods.
For a visual inspection, figure~\ref{fig:network_transport} shows the network of the transport's codes, highlighting the two dis-aggregation paths in blue and red.
Remarkably, only the blue codes sum up to the value of their parents.
Instead, the red nodes do not sum to the available value of the parents, although it is possible that using the hierarchical tree obtained from the codes does not account properly for the red classification.

Contrarily, the alternative classification maintains only a single hierarchical structure removing the double methodologies, and the quality is corroborated by the fact that the sum of the values of the sons return the value of the parent codes.
We want to stress that the scenario of the two methodologies derives from a misleading interpretation of the indices offered by the IMF and not an error in their implementation of the database, since it is possible to reconstruct the correct structure.
\begin{figure}[!t]
	\centering
	\includegraphics[scale=0.5]{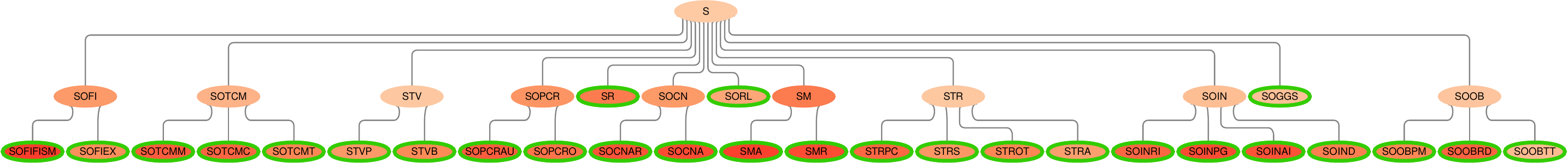}
	\caption{The tree graph of the relation between the BOP codes following the \emph{alternative} classification. The colour of the nodes represents the frequency of missing elements found in the raw database with red corresponding to 1 and white corresponding to 0. The green circle indicates the codes considered in the final construction of the Universal database (see text below).}
	\label{fig:net}
\end{figure}
We present the first three main layers of aggregations of the \emph{summable} classification in figure~\ref{fig:net}.
Among the layers, the first one accounts for the total sum each country produces in a year.
The second and third layers are more representative and diverse of the heterogeneous aggregation of the services.

\subsection*{Missing values in Services data}

In the construction of the Universal database we select the codes of table~\ref{tab:complete_set}, covering with the lower level of aggregation.
For the sake of clarity we highlight these codes in figure~\ref{fig:net}  with a green thick border, and we refer to the set as the \textit{complete set} of codes, since with its knowledge it is possible to obtain the full database.
\begin{table}[h!]
	\centering
	\small
	\begin{tabular}{| l | c | c |}
		\hline
		\textbf{code} & \textbf{layer} & \textbf{description}\\
		\hline
		BXSOGGS & 1 & Government Goods and Services \\
		BXSORL & 1 & Charges for the Intellectual Property \\
		BXSR & 1 & Maintenance and Repair \\
		\hline
		BXSMA & 2 & Manufacturing Services, Goods for Processing Abroad \\
		BXSMR & 2 & Manufacturing Services, Goods for Processing Inside \\
		BXSOTCMT & 2 & Telecommunications Services \\
		BXSOTCMM & 2 & Information Services \\
		BXSOTCMC & 2 & Computer Services \\
		BXSOPCRO & 2 & Cultural and Recreational \\
		BXSOPCRAU & 2 & Audiovisual \\
		BXSOOBTT & 2 & Technical, Trade-related, and Other Business \\
		BXSOOBRD & 2 & Research and Development \\
		BXSOOBPM & 2 & Consulting \\
		BXSOINRI & 2 & Reinsurance \\
		BXSOINPG & 2 & Pension \\
		BXSOIND & 2 & Direct Insurance \\
		BXSOINAI & 2 & Auxiliary Insurance Services \\
		BXSOFIFISM & 2 & FISIM \\
		BXSOFIEX & 2 & Explicitly Charged and Other Financial Services \\
		BXSOCNA & 2 & Construction Abroad \\
		BXSOCNAR & 2 & Construction Inside \\
		BXSTVB & 2 & Travel Business \\
		BXSTVP & 2 & Travel Personal \\
		BXSTRS & 2 & Sea Transport \\
		BXSTRPC & 2 & Postal and Courier \\
		BXSTROT & 2 & Other Passenger Transport \\
		BXSTRA & 2 & Air Transport \\
		\hline
	\end{tabular}
	\caption{The list of the 27 codes implemented in the Universal database with their description. The initial 2 letters `BX' indicate the credits.}
	\label{tab:complete_set}
\end{table}

The BOP data covers about 197 countries although we select the subset of 160 countries intersecting the set of countries available for the Goods, and covering basically all the relevant economies in the world in terms of economic impact.
Finally, in terms of temporal resolution, BOP goes back up to 1940 with very few sectors in the first years.
However, we reconstruct the series starting from 1990 because the Universal database is constraint by the presence of also the Goods database, which starts in 1996.

\begin{figure}
	\centering
	\includegraphics[scale=0.5]{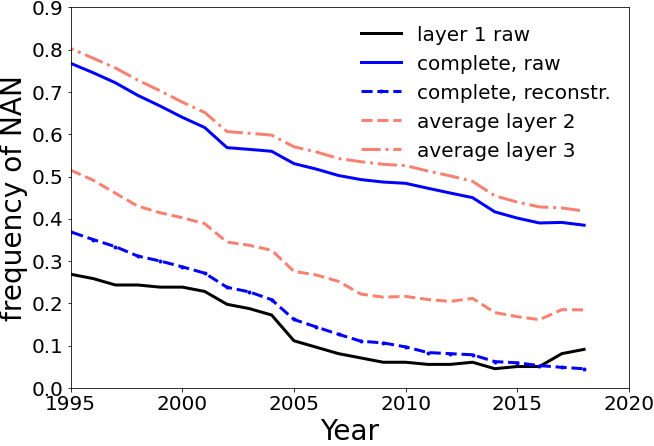}	
	\caption{Proportion of missing elements in the complete set of indices of the BOP database as a function of the years in the layers of the raw database (160 countries).
		The lines with the shade of blue correspond to the different layers of the raw database while the red and dark red lines correspond to the complete set in the raw (red) and reconstructed (dark red) cases.}
	\label{fig:freq_nan}
\end{figure}
Focusing on the temporal distribution of the missing elements, it is clear that the raw database is not uniform, as shown in figure~\ref{fig:freq_nan}.
In the early years of the series there is a larger presence of missing values with a decreasing global trend approaching the recent times.
In the developed economies these behaviours are primarily induced by the complete absence of the first segments of the series for all the indices.
For example, in the case for Austria, no information is available before 2005; regarding Belgium before 2004, and Japan before 2000.
Instead, less developed countries present subsets of the indices with missing scores on short to medium temporal windows, while single NaNs are rarer in the datasets.
Hence, in between 2005 and 2015 the raw database offers its better quality, even if the raw version continues to have a high fraction of missing values.

Finally, also the geographical coverage of the raw database is not uniform, as shown in the top map of figure~\ref{fig:maps_nan}.
Many countries have a large fraction of missing values, even if some of them are developed economies.
Remarkably, Switzerland, China, Great Britain, and Spain are among the developed countries with a low quality representation in the raw data.
This situation of heavy lack of data and heterogeneity calls for a specific intervention to reconstruct the missing elements, which is the subject of the next section.

\subsection*{Reconstruction of the missing Services}

We implement a reconstruction of the BOP database using different machine learning techniques.
The use of the alternative classification discussed above allows the derivation of the full hierarchy reconstructing the complete subset of leafs by summing the indices (red-circled indices in figure~\ref{fig:net}).
However, the knowledge of the values of the upper layers of the classification are used in the reconstruction because we do not want to change abruptly the relative importance of the nations or of the daughter indices.
Moreover, in the following we assume that any not-missing value of the raw database is \textit{correct}, thus also the entries with zero value are labelled as correct.
A very basic and easily interpretable reconstruction technique is given by the linear interpolation of the temporal series in the forward direction, \textit{i.e.} generating the elements for which there exists some previous information.
Therefore, the linear interpolation is assumed as the reference method, giving relatively good results despite the simplicity of the method.
Successively, we consider two families of methodologies: the Random Forest class and the Nearest-Neighbour imputations~\cite{batista2002study} class.

The Random Forest consists in an ensemble of decision trees~\cite{Breiman2001} trained to optimize the entropy of the reconstruction.
Rather, a decision tree is a single and non-overlapping subdivision of the space of the features, \textit{i.e.} the values of the series, into distinct regions following a hierarchical ordering of the feature-selection with a tree structure~\cite{Safavian1991}.
Since the splits of the tree are randomly chosen, it is possible to derive many different decision trees forming a statistical ensemble.
Using the ensemble of trees it is possible to estimate each missing feature of the input database (here the Services) based on the behaviour of the other entries for which those features are specified.
This method is applied on the complete set of series aggregating the entries in time and space (years and nations), the so-called Temporal Random Forest~\cite{Serafini}, such that the information from similar economies is taken into account.
In reference~\cite{Serafini}, this method is found to be superior in terms of the quality of reconstruction with respect to the separate reconstruction of each country series.

The latter method we realize, the k-Nearest-Neighbour (kNN), consists in the generation of the missing features with a weighted network average from other features.
In detail, the network of similitude between the economies, from the prior knowledge of the indices in the upper layers, is built under the assumptions that the reconstruction does not strongly modify the relative importance of the nations.
Accordingly, the kNN imputer~\cite{Troyanskaya2001} links the first $K$ nearest-neighbours elements from the available information on the upper layer.
The resulting network is then applied for the selection of the nearest elements considered in the weighted reconstruction of the missing values.
Finally, the weight is defined as the Euclidean feature-distance in the original space between the selected entries.

\begin{figure}[!h]
	\centering
	\includegraphics[scale=0.5]{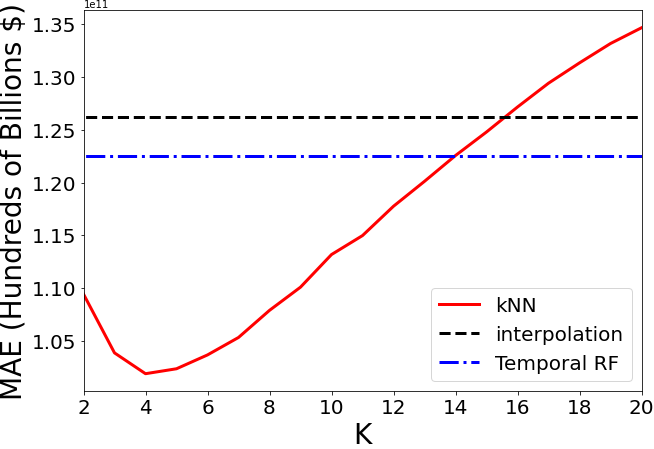}	
	\caption{The Mean Absolute Error of layer of the proposed reconstruction techniques a s function of the number of Nearest-neighbours $K$. }
	\label{fig:mae}
\end{figure}
We test the quality of the reconstructions on a test-set of the database composed by countries and indices for which all the entries are specified.
On this dataset we generate random samples setting some feature as missing, and we generate 100 samples of this process.
For each replica we label as missing at random, and we compute the statistics on the reconstruction techniques described above.
The quality of the reconstructions are finally evaluated by the computation of the Mean Absolute Error (MAE) between the reconstructed and the truth values.
As shown by the left panel of figure~\ref{fig:mae}, the interpolation method down-perform the MAE of the Temporal Random Forest, 
but the best achievement is obtained by using the kNN network built with only a few of neighbours.

\begin{figure}[!h]
	\centering
	\includegraphics[scale=0.125]{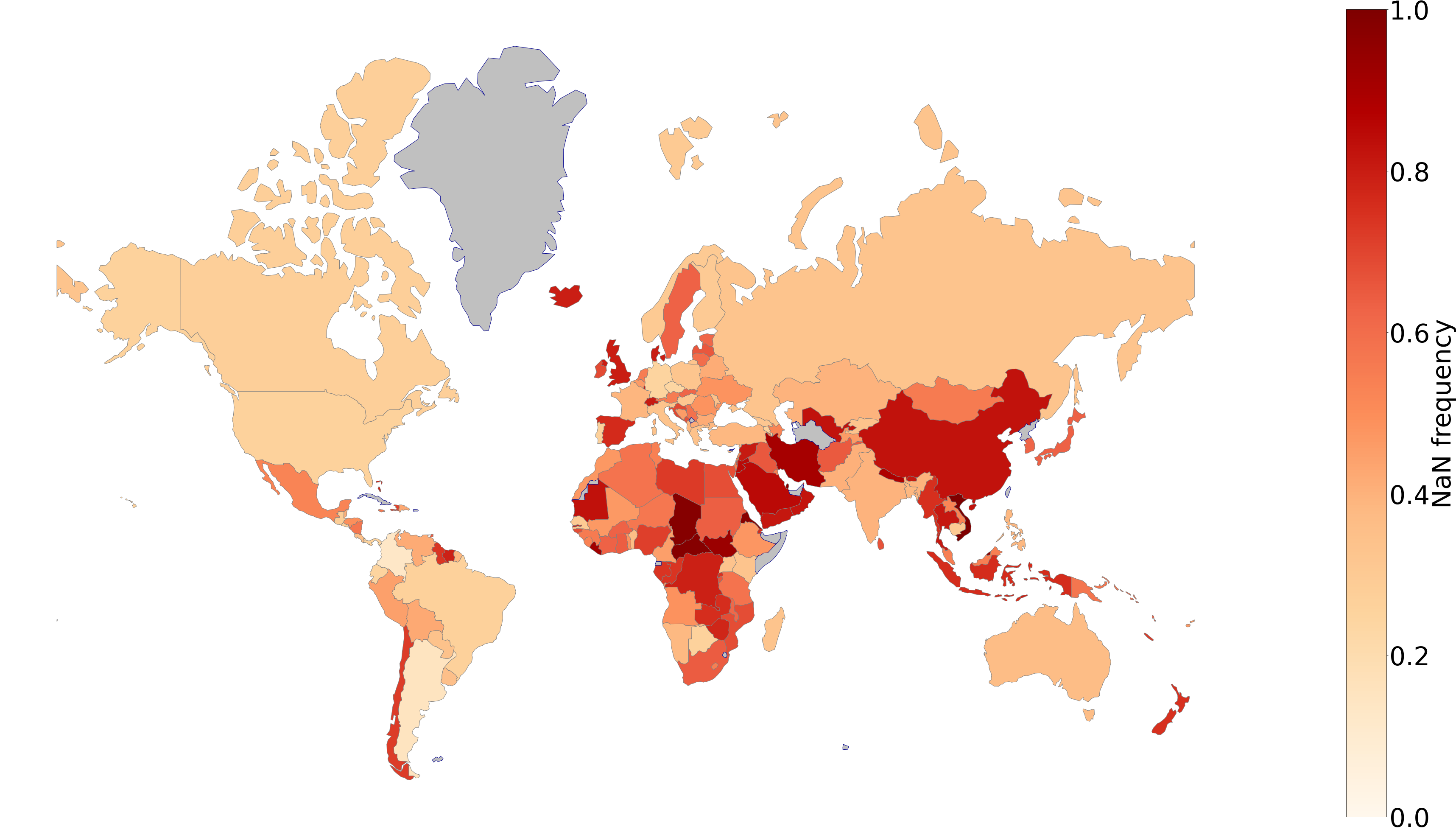}
	\includegraphics[scale=0.125]{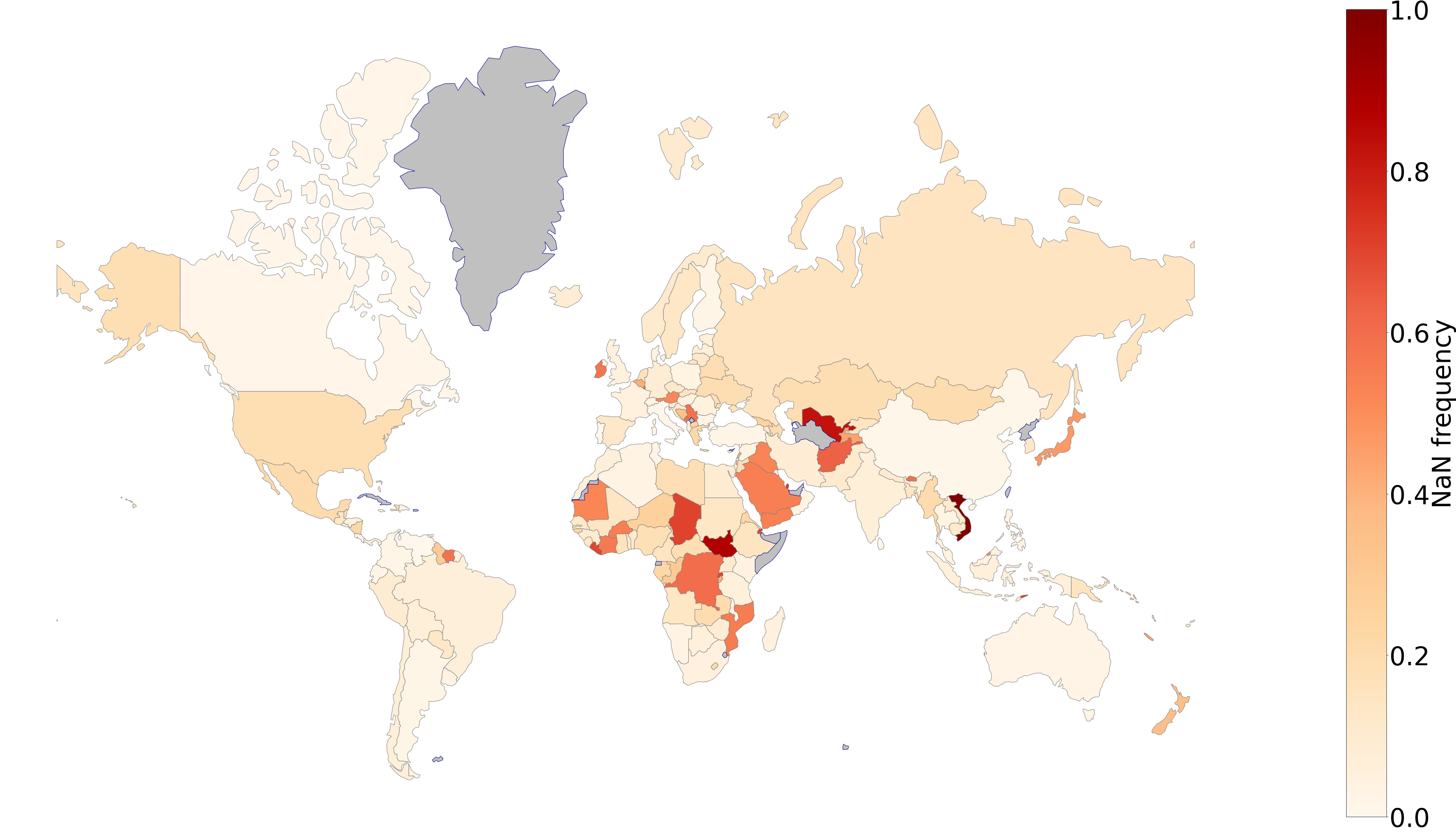}
	\caption{Maps of the World where the colour correspond to the density of missing elements in the complete set of indices of the database of the Services. 
		The top map shows the density in the raw data while the bottom map shows the density in the reconstructed data.}
	\label{fig:maps_nan}
\end{figure}
Therefore, in order to reconstruct the database we will use in the following analyses we implement the kNN reconstruction technique with $K=5$~\footnote{We may expect that size effects require a larger $K$ and the difference in MAE between 4 and 5 in not appreciable.} neighbours on the complete set of indices described in the previous section.
Figure~\ref{fig:maps_nan} shows World maps drawing the colours of the countries in terms of the fraction of missing elements in the raw (upper) and reconstructed (lower) databases and allowing a visual representation of the quality of the reconstruction.
Finally, the reconstructed set, truncated between 1996 and 2018, is then aggregated to the 2-digits UN-COMTRADE data in the so-called \textit{reconstructed Universal database}.

\subsection*{The Economic Fitness and Complexity input matrices}
The various Economic Complexity measures are computed starting from quantities derived from the export data.
The raw export volumes are not directly used essentially for two reasons: i) they trivially depend on both the sizes of the sector and the country and ii) they do not provide an assessment of the \textit{competitiveness} of the given country in exporting a product. 
In order to overcome these limitations, the Revealed Comparative Advantage (RCA), introduced by Balassa~\cite{Balassa1965}, is commonly used in the literature. 
In formula, the RCA of country $c$ in activity $a$ is computed as:
\begin{equation}
	\mathrm{RCA}_{c,a}=\frac {E _{c,a}/\sum_a E _{c,a}}{\sum_c E _{c,a}/\sum_{c,a} E _{c,a}}
\end{equation}
where $E_{c,a}$ is the value of the Product or Service sector $a$ (what we generically call an activity), in constant US dollars, exported by country $c$. 
Note that this formulation permits to identify a natural threshold of 1 to determine whether $c$ exports $a$ in a competitive way or not with respect to what the global set does. 
As a consequence, we can define the binary matrix $\textbf{M}$ whose element $M_{c,a}$ is equal to 1 if $\mathrm{RCA}_{c,a} \ge 1$, and it is equal to 0 otherwise. 

We also define the Marked Share (MS) matrix element as:
\begin{equation}
	\mathrm{MS}_{c,a}=E _{c,a}/\sum_c E _{c,a}
\end{equation}
that is an assessment of the importance of country $c$ in the global trade of $a$. 
Note that in this case no natural threshold is available, and a residual correlation with the size of the country is usually present.
Hence, we will call the Fitness based on $\textbf{MS}$ the Extensive Fitness, recalling the concept of extensive and intensive quantities in statistical physics, and we may indicate the \textit{standard} computations based on the binary RCA as intensive for comparison.

Both the series of the Market Share and of the binary RCA present their sources of noise and different techniques can be implemented in order to reduce their negative effect.
However, for the sake of simplicity and since the data at the aggregated level of 2-digits usually consider large volumes, in both the series of Market Share and RCA we apply a simple exponential smoothing with half-life of 3 years.
The use of the exponential smoothing allows the reduction of the noise has the positive side effect of maintaining a persistence of a few years in the temporal series, without the use of more complex Machine Learning Models.

\section*{Results}

The Economic Fitness and Complexity framework can be divided in two lines of research.
The first one aims at assessing the \textit{Relatedness} between sectors and sectors, and countries and sectors~\cite{hidalgo2018principle}. 
Usually, the output is a network of products, such as the Product Space~\cite{Hidalgo2009} or the Taxonomy Network~\cite{Zaccaria2014}. 
Actually, in this paper we will compute the Product Progression Network \cite{Zaccaria2018services}, based on the Assist Matrix validation framework introduced in~\cite{Pugliese2017}. 
With respect to other approaches, the one used here has two advantages: i) the time evolution is explicitly taken into account, permitting the construction of a \textit{directed} network of sectors, and ii) each link is statistically validated by comparing its weight against a distribution arising from a suitable null model.
The second line of research in Economic Fitness and Complexity consists in an assessment of the global competitiveness of a country, based on the idea that sophisticated capabilities are needed to export complex products.
One example is the Economic Complexity Index~\cite{Hidalgo2009}. 
Instead, in the present work we use the Fitness measure introduced in~\cite{Tacchella2012}, which has a number of advantages from both a practical and a theoretical points of view. 
For a broad discussion on these arguments we direct the reader to the relevant literature~\cite{Albeaik2017a,Gabrielli2017,Albeaik2017b,Pietronero2017}; herein we only emphasize the predictive power of the Fitness approach.
As shown in~\cite{Tacchella2018}, it outperforms the IMF in forecasting the GDP time evolution.

\subsection*{Goods and Services interaction}
A measure of the interaction between two activities can be obtained considering their co-occurrences, that is counting how often such couple of activities is present in the baskets of different nations. 
In order to avoid the size effect induced by the nested nature of the system, where developed countries are competitive in many activities, and some activities are by far more widespread than other, one has to properly normalize the simple co-occurrences. 
Following~\cite{Pugliese2017}, we compute the probability the information flows from one activity to the other along the shortest paths.
The resulting weighted network, projecting the original bipartite structure into the activity layer, is the Assist Matrix
\begin{equation}
	B_{a,a'}(t,\Delta) = \sum_{c} \frac{M_{ca}(t)}{d_a(t)}\frac{M_{ca'}(t+\Delta)}{u_c(t+\Delta)}.
	\label{eq:assist}
\end{equation}
Note that the time evolution of $\textbf{M}$ is explicitly taken into account, so one has a different network of activities for each $t$ and $\Delta$. 
Moreover, this network is almost fully connected, and the presence of spurious links prevents also simple tasks such as a clear visualization of the connectivity. 
In order to filter the links maintaining only the relevant ones, we have to statistically validate each link against a suitable null model.

The implementation of the statistical validation of the network of interaction is pursued by considering an ensemble of random networks preserving few suitable macroscopic constraints~\cite{Saracco2017}. 
We generate the ensemble of randomized bipartite networks related to the empirical case, and we use them in the generation of the ensemble of the random Assist Matrix at fixed $(t,\Delta)$. 
Thus, we seize only those links in the empirical Assist Matrix that are above a percentile threshold, resulting in the statistically validated matrix at fixed $(t,\Delta)$.
Successively, we validate only those links that are statistically significant on all the available years $t$ at a fixed $\Delta$, aiming to reduce the false inference problem.
Therefore, we have the topological graph keeping only the links that are found relevant on all the temporal series analyzed.
Finally, we consider the presence of auto-correlation on the temporal series of the raw signal, and therefore we collapse a range the graphs into a single network where each link has a weight equal to the number of times it has been validated at different $\Delta$.

In detail, in the present work the random ensemble considered for the validation of the network of interaction is the Bipartite Configuration Model~\cite{Squartini2011,Saracco2015} (BiCM).
The BiCM randomizes all the information available but the degrees of the nodes, \textit{a.k.a.} the ubiquity and the diversification, which are kept constant (on average).
Indeed, activities with a larger ubiquity have higher probability to be produced by a random country by chance and the BiCM contemplate this information.
For the link selection we apply a \textit{small} percentile threshold of $95\%$, compared to other cases in literature, because we require such level of statistical relevance along all the time series.

\begin{figure}[!h]
	\centering
	\includegraphics[scale=0.5]{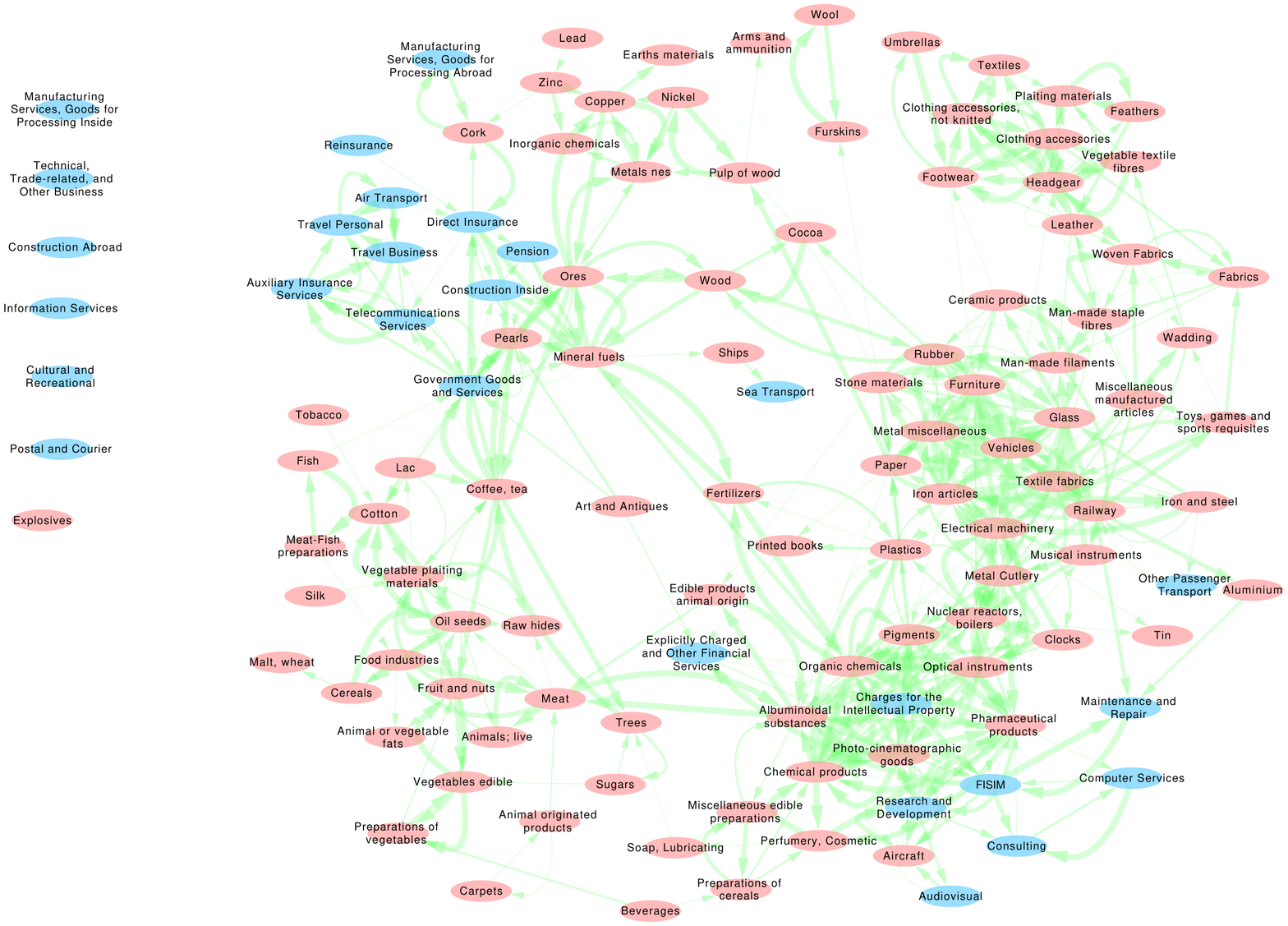}
	\caption{\textbf{The statistically validated network of sectors}. Goods are in pink, while Services are in light blue. The width of the links is proportional to the number of statistically validated time intervals.}
	\label{fig:network_hs}
\end{figure}
The collected network spans 10 years of possible time delays ($\Delta\in[0,10]$) in order to keep track of the autocorrelation.
Figure~\ref{fig:network_hs} shows the network following the HS classification.
On the bottom right, a High-Tech cluster contains both high Complexity Goods and Services. 
Going counter-clockwise, immediately above we find all the Heavy Industries and then Textile related sectors.
Few other clusters collects the mineral Raw Materials and the Vegetable and the living Materials. 
These sectors show a lower degree of connectivity, indicating that countries specialized in these sectors rarely jump to more complex industries.
The Services are not strongly dispersed but are not forming a single block of nodes. 
Note that High-Tech related Services, those associated to royalties and R\&D, are strongly connected with the high Complexity sectors. 
Regarding only the Service, solely the group of transport nodes create a cluster \textit{per se}.
Finally, the statistical nature of the construction of the network allows that few nodes are not connected to any elements because their links are not found to be statistically relevant.
A lower percentile threshold may connect them at the price that the final network of interaction will be less statistically significant with more likely erroneous inferences.

\subsection*{Economic Fitness and Complexity}
The Fitness and Complexity algorithm~\cite{Tacchella2012,Cristelli2013,Pugliese2016} aims at assessing the competitiveness, or \textit{Fitness}, of a country, and the sophistication or \textit{Complexity} of products using a set of coupled equations. 
The idea is the following.
Each country is characterized by its endowments, or capabilities, which represent its social, cultural, and technological structure~\cite{Dosi2000}. 
These capabilities are expressed in what a country produces and exports, so sectors (physical Goods and Services) and their Complexities are linked to the Fitness of each country; in particular, the Complexity of a sector increases with the number and the quality of the capabilities needed in order to be competitive in it, and the Fitness is a measure of the Complexity and the number of the competitively exported sectors. 
In order to make this line of reasoning more quantitative, we start by considering the global structure of the matrix $\textbf{M}$ defined above.
Once countries and products are suitably arranged, the matrix $\textbf{M}$ is triangular, or nested~\cite{Mariani2019}, showing that developed countries have diversified exports, while less developed countries export fewer, lower Complexity products, and these products are actually the ones exported by all countries. 
In order to leverage on this structure to extract information about countries’ competitiveness and sectors’ Complexity, Tacchella et al.~\cite{Tacchella2012} proposed the following set of non-linear, coupled equations
\begin{eqnarray}\label{pilrs1}
	\begin{cases}
		\widetilde{F}_c^{(n)}=\sum_a M_{ca} Q_a^{(n-1)} & \\ \\
		\widetilde{Q}_a^{(n)}=\dfrac{1}{\sum_c M_{ca} \dfrac{1}{F_c^{(n)}}} \\
	\end{cases}
	\begin{cases}
		F_c^{(n)}=\dfrac{\widetilde{F}_c^{(n)}}{<\widetilde{F}_c^{(n)}>_c}  & \\ \\
		Q_a^{(n)}=\dfrac{\widetilde{Q}_a^{(n)}}{<\widetilde{Q}_a^{(n)}>_a}\\ 
	\end{cases}
	\label{eq:f-q}.
\end{eqnarray}
Here $<\cdot>_x$ denotes the arithmetic mean with respect to the possible values assumed by the variable dependent on $x$, with initial condition:
\begin{equation}\label{pilrs2}
	\sum_a Q_a^{(0)}=1   \hspace{6pt}\forall \, a.
\end{equation}

The iteration of the coupled equations leads to a single fixed point that does not depend on the initial conditions~\cite{Tacchella2012,Cristelli2013,Pugliese2016}.
The fixed point defines the non-monetary metrics quantifying the Fitness $F_c$ and the Complexity $Q_a$. 
The convergence properties of equations~\eqref{eq:f-q} are not trivial and have been extensively studied by Pugliese~\textit{et al.}~\cite{Pugliese2016}.
The coupled equations in~\eqref{eq:f-q} relies only on one input, the matrix $\textbf{M}$ stating which countries are competitive in which sectors. 
Usually, the matrix is represented by the binary RCA, although also the Market Shares can be entered in equation~\eqref{eq:f-q}, return the Extensive Fitness~\cite{Tacchella2012,Zaccaria2018services}.
The immediate consequence of this choice is a higher correlation with the size of the country, as expressed for instance by its GDP.
\begin{figure}[!h]
	\centering
	\includegraphics[scale=0.33]{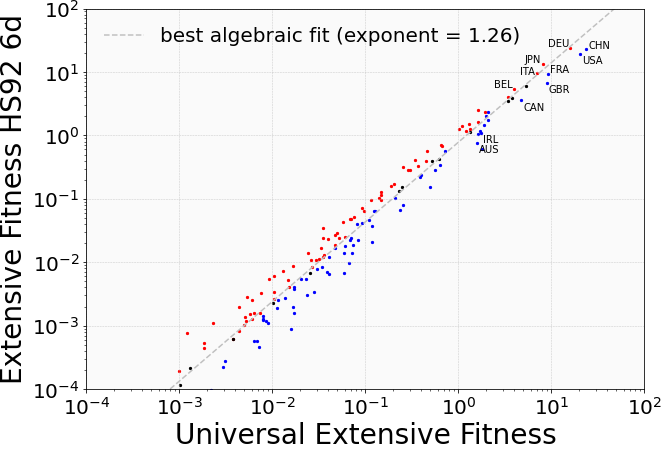}
	\includegraphics[scale=0.33]{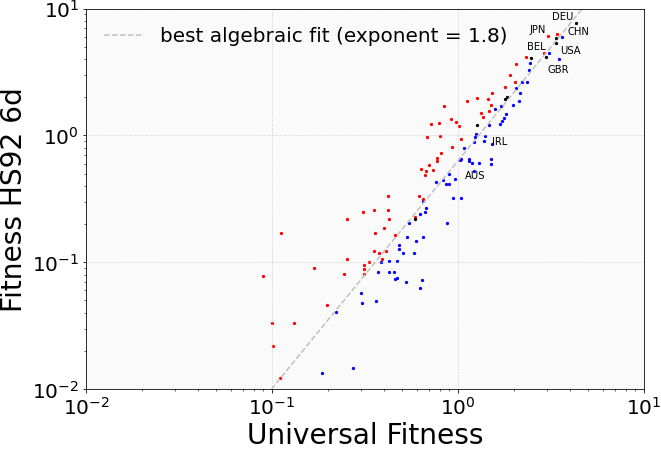}
	\caption{\textbf{6d vs 2d and with/without services comparison}. One the left, the effect on the Extensive Fitness values of integrating services and aggregating into 2 digit sectors. On the right, the same analysis performed using the intensive Fitness. In both cases a good correlation is present, and the major deviations are given by services and good driven Economies. The dashed lines indicate the least square regression fit with a power law shape, while the colour scheme highlight the countries gaining ranking positions in ref and losing position in blue.}
	\label{fig:scatters}
\end{figure}
In Figure~\ref{fig:scatters} we compare the results of the Fitness algorithm when different input matrices are used. 
On the left we compare the Universal (i.e., considering Goods and Services) Extensive (i.e., the Market Shares matrix $\textbf{MS}$ is used) Fitness, on the x-axis, with the Extensive Fitness computed at 6-digits, that is, considering only the Goods and the lower available level of aggregation, containing about 4500 different codes. 
On the right, instead, we show the intensive counterpart, that uses as input the binary RCA. 
By visually inspecting the figure one can notice various properties. 
First, a rather good correlation is present, highlighting that the overall economical results based on the Fitness metrics are robust when Services are integrated and the sectors are aggregated from the 6 to 2 digits levels.
More importantly, the positive effect of specializing in high Complexity Services clearly emerges. 
Countries such as Great Britain (GBR) and Germany (DEU) provide two examples of Services or manufacturing driven countries, as correctly reflected by the Fitness indicator.
\begin{figure}[!h]
	\centering
	\includegraphics[scale=0.3]{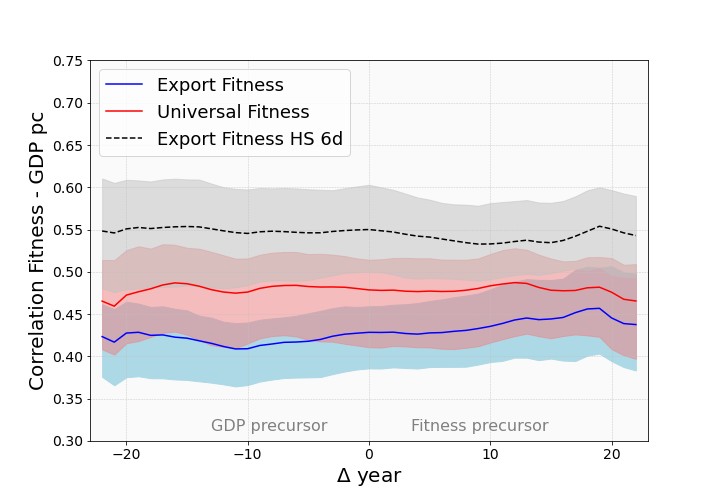}
	\includegraphics[scale=0.3]{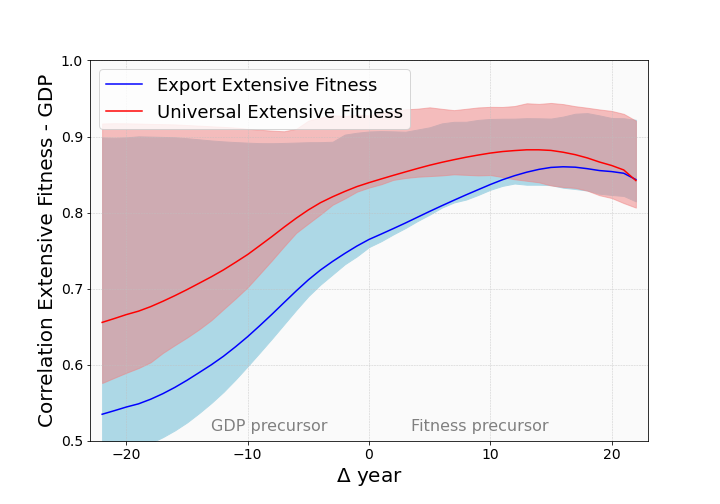}
	\caption{\textbf{Extensive fitness as GDP precursor}. Average correlations (Pearson coefficient) among the GDP and Fitness series with temporal lag. The left panel has the intensive measures while the right panel has the extensive ones. The filled regions is the region in between the $25\%$ and the $75\%$ quantile from the application of the bootstrap. The diagonal line is reported for the visualization of the variation of the strength relations.}
	\label{fig:correlation}
\end{figure}
As shown in references~\cite{Tacchella2018,Cristelli2017}, the Fitness metrics can be used to forecast the GDP variations with a relatively long time interval and high accuracy.
Hence, in this work we address a basic analysis in which we consider the time delayed cross-correlation between the Fitness measures and the respective GDP values with the aim to evaluate if the aggregation of the Services possibly introduce more signal or more noise in the forecasting.
In Figure~\ref{fig:correlation}, on the left, we plot the correlation between the intensive Fitness and the GDP per capita (GDP pc). 
One can easily see that lines are essentially flat, indicating that, albeit correlated, there is no clear time direction from Fitness to GDP pc or vice-versa for either measure.
Interestingly, this is the case of also the 6-digits Fitness usually implemented in the GDP analysis.
On the right, we show the extensive counterpart: the Fitness computed using the Market Shares and the total GDP, not normalized using the population.
The Extensive Fitness, and in particular the Universal Extensive Fitness, that contains both physical Goods and Services, is clearly able to predict the GDP with a time delay of up to 20 years, that however is of the order of the maximal extension of our data. 
Note that we are using the term prediction in the Economic sense, \textit{i.e.} this not an out of sample forecast but a time delayed correlation. 
In this sense, we can safety say that the Universal Extensive Fitness is able to statistically anticipate, or is a precursor of the total GDP. 
However, the converse is not completely true: although there is a signal of correlation, the GDP has a lower predictive power and the Fitness is somehow harder to be anticipated.\\ 
\begin{figure}[!h]
	\centering
	\includegraphics[scale=0.35]{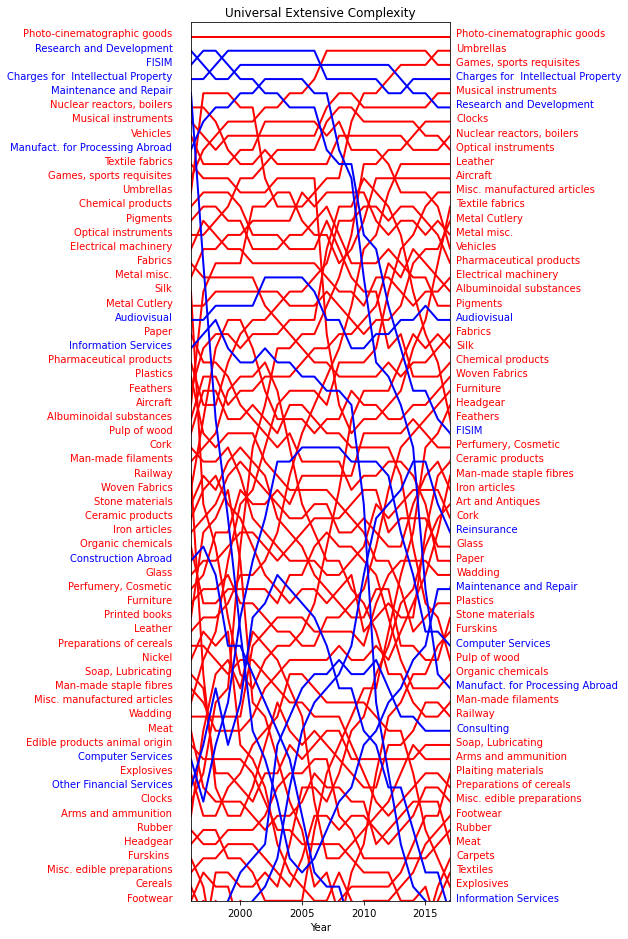}
	\includegraphics[scale=0.35]{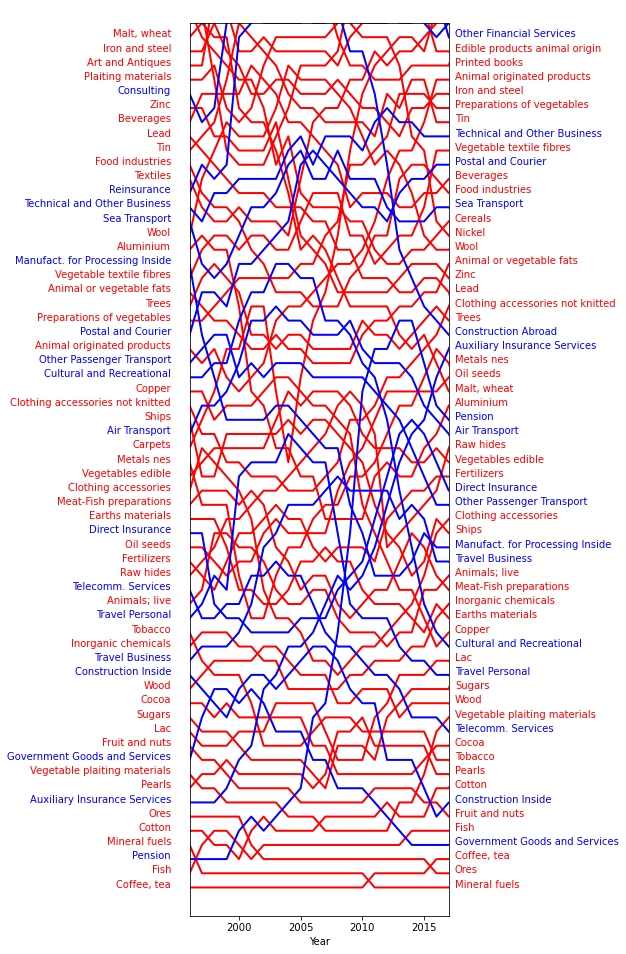}
	\caption{The series of the ordering of the Universal Sectors. The red colors refer to the physical Goods while the blue colors indicates the Services.}
	\label{fig:complexity_orders}
\end{figure}
Finally, we show in figure~\ref{fig:complexity_orders} the time evolution of the Complexity rankings for both Goods (in red) and Services (in blue). 
Even if some noise is present, the rankings clearly follows the layman's intuition of which sector may be sophisticated and which is not. 
In the high Complexity rankings we find services such as R\&D and royalties, and manufacturing sectors such as nuclear reactors, optical instruments, and aircraft. 
Instead, low Complexity products correctly correspond to lower capabilities requirements such as Agrifood sectors.
Note that some top code such as 'Games, sport requisites' includes also accessories and sophisticated products necessary for any kind of sport and competition, while others codes like Umbrellas are boosted by the fact that more than $83\%$ of their Trade is performed by China.
%

\section*{Discussion}
All Economic Complexity measures are based on countries-activities bipartite networks, where the word \textit{activity} usually indicates a physical product. 
Nonetheless, a fast growing fraction of both the International Trades and local economies is based on Services.
Recent attempts tried to integrate the standard UN-COMTRADE database of physical Goods with the IMF BOP database collecting the Service data~\cite{Stojkoski2016,Zaccaria2018services,Mishra2020}.
However, two problems prevented the necessary achievements to reach a level of data control and sanitation comparable to the Goods-only data.
Firstly, the IMF classification is rather convoluted and, as it is, not suitable for a direct application into the EC framework, and the main issue is related to the structure of the classification tree.
Secondly, the significant presence of missing values at different scales of aggregation prevents the realization of a satisfactory level of detail of the classification, forcing the use of various gimmicks to arrange the data.
For instance, Stojkoski et al.~\cite{Stojkoski2016} modified the original Fitness and Complexity algorithm to make it converge. 
In this work we address and propose a solution to both these problems by selecting, on one hand, a self-consistent and not overlapping Service classification and, on the other hand, filling the missing values with a suitable machine learning approach. 
In particular, we compare and test different sanitation strategies and find that a kNN approach gives the lowest reconstruction error.

Once the reconstruction is performed, the \textit{Universal} dataset is available integrated, Goods and Services.
Hence, we applied the new dataset in the Economic Fitness and Complexity framework.
In particular, we build a statistically validated network of Universal sectors bringing to light the interaction among them. 
By construction the interaction is directed because the time evolution of the input matrices is explicitly taken into account.
The resulting network is filtered by validating the interaction with the use of an ensemble of random matrices preserving the degree sequences of nations and activities. 
Thus, only the statistically significant connections are maintained.
Meaningful patterns indicate that there exist a deep interrelation between High-Tech manufacturing and complex Services, and there is a partial isolation of the agrifood sector.

Successively, we computed the Fitness of countries and the Complexity of universal sectors, both in the extensive and the intensive fashions. 
The importance of Services once again emerges, as previous measures of Fitness clearly underestimate the competitiveness of countries such as the United Kingdom. 
Finally, we considered the so-called Extensive Fitness that, being computed starting from the Market Shares, is more connected with the cumulative importance of the country at the International level. 
We show that the Extensive Universal Fitness is a precursor of the GDP, in the statistical sense that it is highly correlated with the GDP when a time delay is accounted. 
In particular, this correlation increases with time, showing that the Universal Extensive Fitness anticipates the GDP, significantly more than the contrary.
This work and most importantly the Universal database, opens up a number of possible lines of research. 
Firstly, even if the results of this paper are focused on the Economic Fitness and Complexity framework, the reconstruction of the Services database and its integration with the Goods data is of general interest and can be also used with other macroeconomic purposes. Secondly, the Universal Fitness can be used to build GDP and GDPpc forecasts using the Selective Predictability Scheme, as done in~\cite{Cristelli2017,Tacchella2018}. 
We expect that the inclusion of Services will improve the predictive performance of this approach. 
Finally, the Universal Progression Network, with or without the support of machine learning approaches, can be used to predict the future activation of sectors now absent in the export basket of countries.


\section*{Data Availability}
The datasets generated and analysed during the current study are available in the \emph{Universal database} repository, \url{ https://efcdata.cref.it/}.

\section*{Usage Notes}
The raw database of the Services is the Balance of Payments, collected by the International Monetary Fund (IMF), which is free of charge, as described in the terms of the \emph{Copyright and Usage} (https://www.imf.org/external/terms.htm).
All the database and the files containing the results of the Fitness and Complexity algorithm are available in the \emph{csv} (comma-separated-values) format of Unicode (UTF-8) string format.
The format of the file requires that the first column contains the row indexes while the first row contains the column indexes.
The choice of the format of the database allows the implementation using various open-source codes and platform, as well as Office suites (and open or free versions).

\section*{Code availability}
The repository contains the Jupyter notebooks written in Python 3 necessary to reproduce the reconstruction and the Fitness and Complexity algorithm.

\section*{Acknowledgements}
We gratefully acknowledge funding received from the Joint Research Centre of Seville (grant number 938036-2019 IT).

\bibliographystyle{unsrt}
\bibliography{biblio}

\end{document}